\documentclass[information,article,accept,pdftex,moreauthors]{Definitions/mdpi} 

\firstpage{1} 
\pubvolume{1}
\issuenum{1}
\articlenumber{0}
\pubyear{2025}
\copyrightyear{2025}
\externaleditor{Marco Leo, Ivan Miguel Pires, Eftim Zdravevski and Petre Lameski} 
\datereceived{2 May 2025} 
\daterevised{7 August 2025} 
\dateaccepted{9 August 2025} 
\datepublished{ } 
\hreflink{https://doi.org/} 

\usepackage{graphicx}
\usepackage{xcolor}
\usepackage{amsmath}
\usepackage{makecell}
\usepackage{url}

\Title{Convolutional 
 Autoencoders for Data Compression and Anomaly Detection in Small Satellite Technologies}
\TitleCitation{Convolutional Autoencoders for Data Compression and Anomaly Detection in Small Satellite Technologies}

\Author{Dishanand Jayeprokash 
 $^{1,}$* 
 and Julia Gonski $^{2}$} 
\AuthorNames{Dishanand Jayeprokash and Julia Gonski}
  \AuthorCitation{Jayeprokash, D.; Gonski, J.}

\address{%
$^{1}$ \quad African Institute for Mathematical Sciences,
    6 Melrose Road, Muizenberg, Cape Town 7945, South Africa\\
$^{2}$ \quad SLAC National Accelerator Laboratory,
    2575 Sand Hill Road MS 95,
    Menlo Park, CA 94025, USA; jgonski@slac.stanford.edu 

}

\corres{Correspondence: jay@aims.ac.za}

\abstract{
Small satellite technologies have enhanced the potential and feasibility of geodesic missions through the simplification of design and decreased costs allowing for more frequent launches. 
On-satellite data acquisition systems can benefit from the implementation of machine learning (ML) for better performance and greater efficiency on tasks such as image processing or feature extraction.
This work presents convolutional autoencoders for implementation on the payload of small satellites, designed to achieve the dual functionality of data compression for more efficient off-satellite transmission and at-source anomaly detection to inform satellite data-taking. 
This capability is demonstrated for the use case of disaster monitoring using aerial image datasets of the African continent, offering avenues for both the implementation of novel ML-based approaches in small satellite applications and the expansion of space technology and artificial intelligence in Africa.}


\begin{document}
\section{Introduction}
\label{sec:intro}

Small satellites are artificial satellites characterized by weights less than 500 kg and placed in orbit around planets and natural satellites to collect information or for communication~\cite{KOPACZ2020}. 
They are used to make Earth observations (EOs) across academia and industry, with relevance in a variety of fields such as basic research and technology demonstration~\cite{Fevgas2025}, environmental monitoring~\cite{ALZUBAIRI2024e02391}, and disaster management~\cite{BATTISTINI2022231}.
In recent years, the development of low-Earth-orbit satellite constellations and the democratization of space have boosted the launch of small satellites. 
This trend is expected to increase in the future~\cite{BARATO2024, KULU2024}.

Due to their remote location and inaccessibility, small satellites face significant constraints in terms of on-board storage, processing power, and downlink bandwidth. 
This has led to advances in handling large EO data, such as the development of new technologies to store and downlink data to the ground, as well as advanced on-board satellite processing operations \cite{LI2021, CHINTALAPATI2024}. 
Artificial intelligence (AI) systems have proven to be excellent in extracting information from a high volume of data, such as images and videos, using computer vision and other machine learning techniques~\cite{Krizhevsky2012}. 
The edge deployment of neural algorithms (i.e., at the source of the data) reduces the need to transmit large volumes of raw data, allowing for more efficient bandwidth usage and faster downstream processing.
It also enables the prompt extraction of actionable insights from data streams, supporting faster and more autonomous system responses.

Accordingly, there is a growing interest in exploring AI methods on board to improve small satellite (SmallSat) technologies~\cite{lofqvist2020, CHEN2025104106}. 
$\Phi$-Sat-1 demonstrated the significant role of AI in data analysis and decision-making processes using object detection, which could contribute to environmental monitoring~\cite{GIUFFRIDA2021}. 
Additionally, lossy image compression for SmallSats is a critical element in designing an off-satellite data transmission system, and, thus, it has been broadly explored in the literature~\cite{9172536, rs11070759, rs13030447, 9690871}.

A less explored avenue of image processing for SmallSats is the use of anomaly detection, which can allow for the recognition of anomalous images that could be useful for data analysis and satellite operations.
Anomaly detection is defined as the recognition of unusual elements in a dataset based on a learned description of a data-driven background-only model, commonly provided by machine learning tools. 
For example, anomaly detection can help in the early detection and containment of wildfires, flooding, and other natural disasters~\cite{AGUESPASZKOWSKY2020}.
Anomaly detection can also be used for the rapid detection of instrumental issues, namely, pixel failures in the camera or other disturbances~\cite{10285414}.
Such methods have been studied for satellite telemetry~\cite{natureAD, telemetry2, timeSeries}. 
However, the area of anomaly detection in SmallSat image analysis remains relatively unexplored.

This work studies the dual use of autoencoders for data compression and anomaly detection via on-satellite deployment for SmallSat technologies. 
An autoencoder (AE) is an example of an ML architecture that can offer both functionalities.
Autoencoders are trained to reconstruct inputs after dimensional reduction through a bottleneck latent space, therein providing a lower-dimensional basis for the input that can be used to generate authentic reproductions of the original input after sampling from the compressed representation. 
The use of AEs for data compression has been explored in recent years in the context of SmallSat missions~\cite{9883256}.
The specific model used here is a convolutional AE (CAE); convolutional layers aid the AE in learning from two-dimensional images. 
These results expand on those in the previous literature by providing a proof of concept that the ML models commonly used for intelligent image compression are also able to provide anomaly detection capabilities without the need for further optimization or additional processing resources. 

\begin{figure}[H]
   
   \hspace{-12pt} \includegraphics[width=0.8\linewidth]{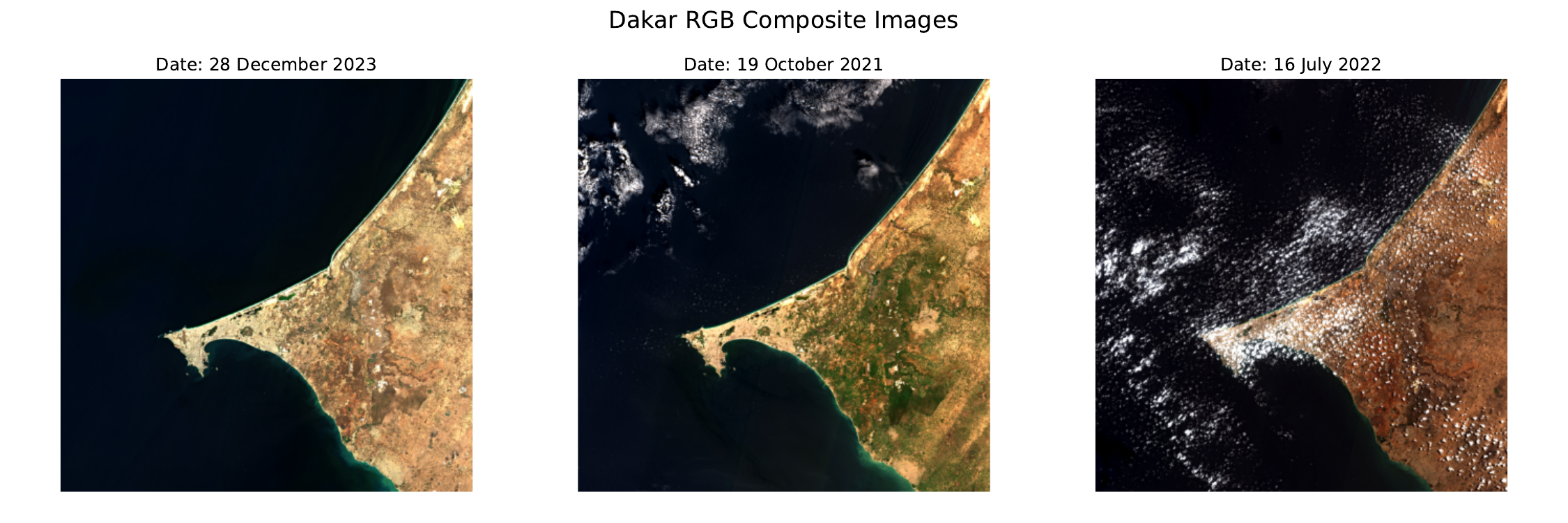}
    \caption{Sentinel-2 
 images of an approximately 110 km $\times$ 110 km aerial region over Dakar, Senegal, used as input to the machine learning models for inference~\cite{dataspace2024}. Three different images are shown with different cloud coverage percentages, with their acquisition dates provided in the title.
    \label{fig:dakar}} 
\end{figure}

Small satellites are cost-effective and can be developed in relatively short timescales compared to other scientific instruments. 
In this way, they offer a testbed for advanced technology by providing new opportunities for payload designs~\cite{SELVA2012}, enabling emerging nations such as Africa to develop and launch space programs.
These technologies can benefit Africa in terms of socioeconomic and environmental development, for example, through their application in agricultural surveillance, real-time disaster monitoring, and sustainable development.
Therefore, the use case of anomaly detection considered here is the identification of Saharan dust storms and camera defects in aerial satellite images of Dakar, Senegal, taken during the Copernicus Sentinel-2 mission (Figure~\ref{fig:dakar}).

\vspace{-4pt}

The ability to compress images to lower-dimensional representations on a satellite could allow for reduced data transmission to Earth, followed by decompression, enabling more power-efficient data downlinks in the highly constrained satellite computing system~\cite{giuliano2024}.
Further, real-time dust storm recognition at the satellite level could allow for dynamic changes in satellite image-taking (i.e., the disruption of standard image-taking patterns to focus on anomalous regions), with significant impacts of improved weather and climate monitoring in the region~\cite{O'Sullivan2020}.


\section{Datasets and Image Processing}
\label{sec:data}

The Sentinel-2 mission is part of the European Union's Copernicus program consisting of a wide-swath, high-resolution, multi-spectral imaging payload~\cite{rs12142291}. The mission consists of twin satellites phased at 180$^{\circ}$, with a high revisit frequency of 5 days at the equator. The payload is a multi-spectral instrument (MSI) that samples 13 spectral bands at different spatial resolutions of 10 m, 20 m, and 60 m. The orbital width is 290 km.

The Sentinel-2 datasets are open-source and easily accessible through the Copernicus Data Space Ecosystem platform. This platform provides a large amount of georeferenced imagery to researchers, developers, and policymakers worldwide ~\cite{dataspace2024}. The dataset includes Level-1C products (orthorectified top-of-atmosphere reflectances) and Level-2A products (bottom-of-atmosphere reflectances). The Sentinel-2 MSI acquires measurements with a radiometric resolution of 12 bits. The same measurements are converted to reflectance and stored as 16-bit integers.
The choice of Sentinel-2 datasets for this study is not only based on their free, open-access policy and large availability of satellite images but also because the data acquired by the MSI on board the satellite is similar to those acquired by the multi-spectral camera used in SmallSats. 

\textls[-15]{The dataset used in this study consists of 70 Sentinel-2 aerial images of the region of Dakar, Senegal, each covering an area of approximately 12115 km$^2$.
Dakar is chosen as the region of interest in Africa because it shows a balance between ocean and land for realistic image feature presence, non-negligible rates of image anomalies such as Saharan dust storm events, and favorable weather conditions for satellite observations (e.g., relatively low cloud coverage).  
A set of images taken from 2020 to 2024 with a 60 m resolution and $<$20\% cloud cover at the L2A processing level is chosen to develop the CAE model, 
comprising 70 images in total. 
Of these, 63 are used for the training set, and the remaining 7 compose the validation and test sets. 
The anomalous dataset consists of two known dust storm events on 1--5 June 2022~\cite{duststorm}. 
These two anomalous Sentinel-2 acquisitions are not included in the 70 images in the non-anomalous training/testing dataset and are reserved for a dedicated test set used to examine the anomaly detection capabilities of the ML approach. }



Before the training dataset was passed to the network, it was preprocessed as follows: (1) bands were selected using free, open-source QGIS (Quantum Geographic Information System) v3.40 software; 
 (2) the dataset was split into training, validation, and test sets; (3) patches (tiles) were extracted with overlap; (4) the patches were saved with their respective metadata; and (5) normalization and datatype conversion were performed. The details of each step are provided below. 

\begin{figure}[H]
 
   \hspace{-7pt} \includegraphics[width=1\linewidth]{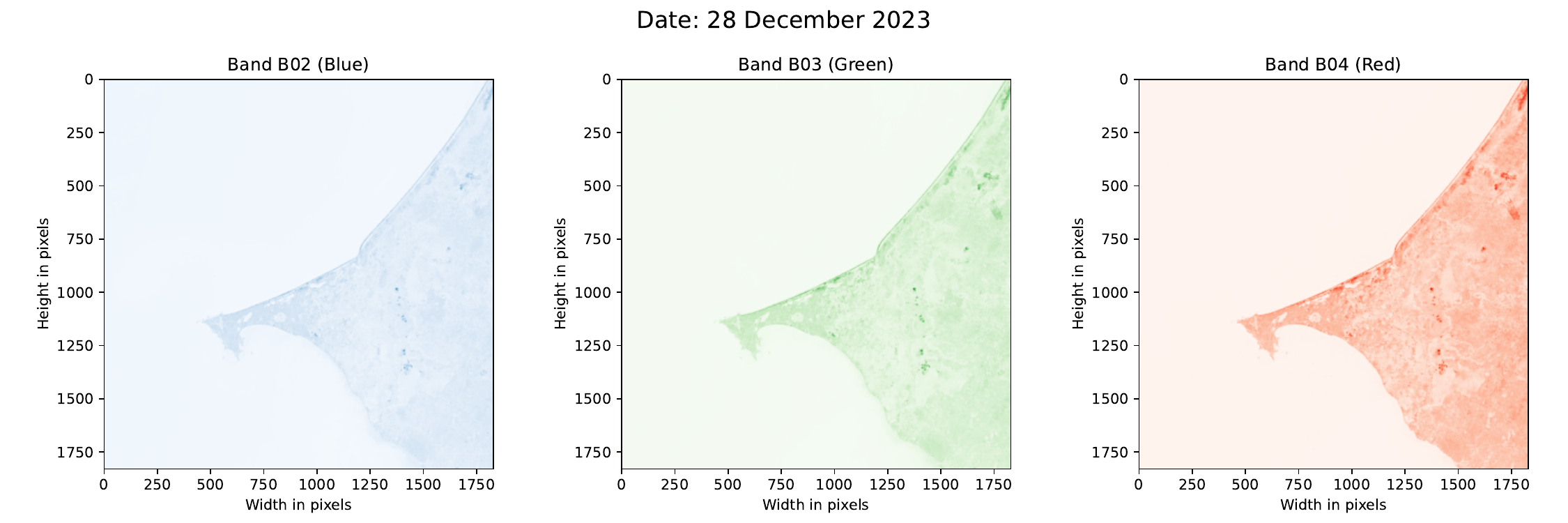}
    \caption{Example Sentinel-2 image split into the individual color bands of input images generated during preprocessing and used for training.
    \label{fig:separated_bands}} 
\end{figure}

Band selection was performed using QGIS. 
Bands 2 (blue), 3 (green), and 4 (red) were selected to isolate them in RGB from the rest, and a composite image was created for visualization, as shown in Figure~\ref{fig:separated_bands}. 
Ultimately, only Band 4 (red) was used to train the CAE  to allow it to properly learn both water and land features, which are predominantly present in the blue and green bands, respectively. 

\textls[-15]{Images were split into training (90\%) and validation/test (10\%) sets based on their date of acquisition.
Each image with a 60 m resolution had a size of 1830 $\times$ 1830 pixels. 
Single-band images were then divided into patches of size $256 \times 256$ pixels, with an overlap of 135 pixels between adjacent patches, achieving 99.89\% image coverage. 
This was achieved by sliding a window across the image with a stride equal to the patch size minus the overlap, ensuring that neighboring patches shared common regions. 
The patch size was chosen because it provides a balance between power consumption and processing time and accuracy~\cite{GUERRISI2023}. 
The overlap was selected to balance comprehensive feature capture with GPU memory constraints, critical for small satellite applications, based on a systematic evaluation balancing image coverage and computational efficiency.
At each position, a patch was extracted and passed to a normalization routine to ensure that the pixel values fell within the range of [0, 1].
Patch extraction was performed after dataset splitting, with identical processing applied to all images in order to prevent data leakage and ensure consistent patch views across sets.}

\vspace{-8pt}


In summary, before preprocessing, the training set has 63 images, the validation has \mbox{4 images}, the test set has 3 images, and the anomaly test set has 2 images.
After preprocessing, the training dataset consists of 12,348 patches, the validation dataset consists of 784 patches, the test dataset consists of 588 patches, and the anomaly test dataset consists of 98 patches, all in Band 4 representation. 
As all images cover the same aerial area over Dakar, patch extraction is applied after the split into training, test, and validation sets, and the patch extraction process is exactly the same for all images. It is ensured that the same image views are passed to the model in both training and evaluation.


\section{Convolutional Autoencoder Model}
\label{sec:CAE}

The architecture of the CAE model is based on the convolutional autoencoder (CAE) proposed in Ref.~\cite{GUERRISI2023}. 
It consists of an encoder with convolutional layers that compresses the image to a lower-dimensional latent space and a decoder with deconvolutional layers that reconstructs the image from the latent space to the original input dimensionality. 
The compression ratio (CR) of the model is shown in Equation~\eqref{eq:CR} and defined as the ratio of the input image size in terms of pixels and the bottleneck size in terms of units:

\begin{equation}
    \text{CR} =  \frac{\text{patch height} \times \text{patch width} \times \text{image depth}}{\text{latent space dimension}}
    \label{eq:CR}
\end{equation}





\textls[-19]{The encoder accepts images of input $256 \times 256 \times 1$ and contains the following: (1) 10 convolutional layers with an increasing number of filters (16, 32, 64, and 128); 
(2) the leaky ReLU activation function and $\alpha = 0.01$ for the better handling of low-intensity features, with each layer having a kernel size of $3 \times 3$ and an alternating stride of 1 and 2 for the downsampling convolutional layer; (3) flattening of the output of the last convolutional layer to convert the 3D tensor into a 1D vector; and (4) a bottleneck layer outputting a 1D vector of 8192-dimensional latent space. 
Therefore, the CAE used here has a latent space with a CR of 8. }


The decoder, which is a symmetrical representation of the encoder, performs the reverse process by returning the reconstructed image from the latent space as the final output. 
It is composed of the following: (1) a latent vector (8192 units) that is reshaped into a 3D tensor of size $8 \times 8 \times 128$; (2) 10 transposed convolutional layers with leaky ReLU activation using $\alpha=0.01$ to upsample spatial dimensions, returning a feature map of size $256 \times 256 \times 16$; and (3) a final layer with a linear activation that generates the output with the original size of $256 \times 256 \times 1$. 
The loss function employed to train the model uses the structural similarity index measure (SSIM)~\cite{Dosselmann2011, HORE2010}, which is designed to align with the human visual perception of images, including image qualities like luminance, contrast, and structural similarities.
For this reason, the SSIM is chosen over the mean squared error loss, which focuses only on pixel-wise differences and may produce blurry reconstructions and, as a result, can compromise an image's structural fidelity.
The SSIM loss is thus defined in Equation~\eqref{eq:SSIM_loss}, where $y_{true}$ and $y_{pred}$ are the true and reconstructed images, respectively. $y_{true\hspace{0.1cm}i}$ and $y_{pred\hspace{0.1cm}i}$ represent the image contents in the $i$th local window, and $M$ represents the number of local windows in \mbox{the patch}.

\begin{equation}
    \label{eq:SSIM_loss}
    \text{SSIM Loss} (y_{true}, y_{pred}) = 1 - \frac{1}{M} \sum_{i = 1}^{M}\text{SSIM}(y_{true \hspace{0.1cm}i}, y_{pred \hspace{0.1cm}i}). 
\end{equation}

\begin{figure}[H]
   
  \hspace{-14pt}  \includegraphics[width=1\linewidth]{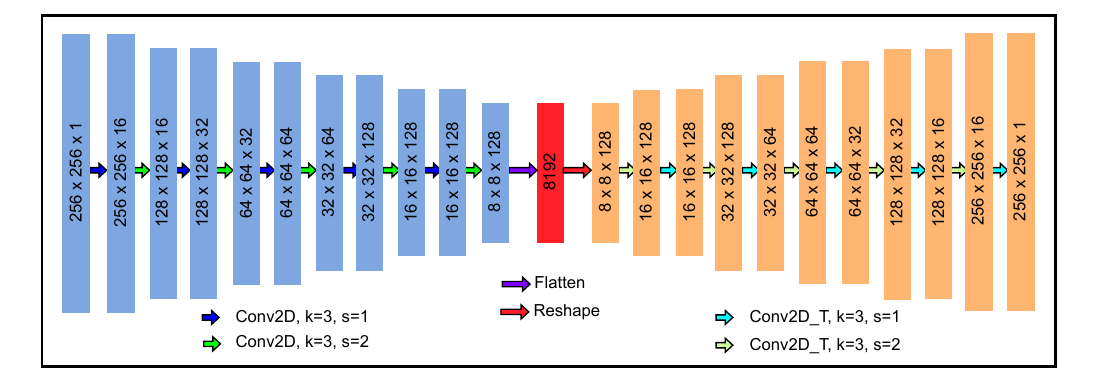}
    \caption{CAE architecture 
 used in this study. \textls[-15]{The encoder is shown in blue, the bottleneck is shown in red, and the decoder is shown in orange. The shape of each block is in the format of height $\times$ width $\times$ depth. The arrows represent the operations in each layer with the following details: Conv2D indicates convolutional layers; $k$ is the kernel size; $s$ is the stride; and Conv2D\_T indicates the transposed convolutional layers.}}
    \label{fig:CAE}
\end{figure}

Figure~\ref{fig:CAE} provides a diagram of the CAE used in this study. 
It is trained over the preprocessed input image set described in Section~\ref{sec:data} for 250 epochs with a batch size of 8, using the Adam optimizer~\cite{kingma2017adammethodstochasticoptimization} with a learning rate of 10$^{-4}$.
As this work seeks to emulate the state-of-the-art CAE described in Ref.~\cite{GUERRISI2023} and demonstrate the added anomaly detection capability, further optimization of hyperparameters is not performed in \mbox{this scope}. 
\vspace{-15pt} 

\section{Results}
\label{sec:result}

After training, the CAE is able to perform an intelligent compression of the images taken by the small satellite to a dimensionality with a compression ratio of 8. 
The CAE output SSIM score for each event can be used to describe the fidelity of the reconstructed output, with high SSIM scores indicating higher-quality reconstructions.
Furthermore, events with an exceptionally poor reconstruction quality (or lower SSIM scores) are more likely to be anomalous with respect to the training set. 
In this way, the SSIM can also be used as a classifier for anomalous images. 
The performances of the CAE latent space and SSIM score in both compressing data and identifying anomalies are presented below. 

\subsection{Image Reconstruction and Compression}

With a compression ratio in the latent space of 8, the proposed transmission of image data off-satellite using the CAE latent space encoding results in a reduction in the off-satellite data rate of 12.5\%. 
Such data rate reductions can allow for lower-power data transmission systems, which are especially valuable in the power-limited environment of satellites.

\begin{figure}[H]

\begin{adjustwidth}{-\extralength}{0cm}
\centering 
 \includegraphics[width=0.99\linewidth]{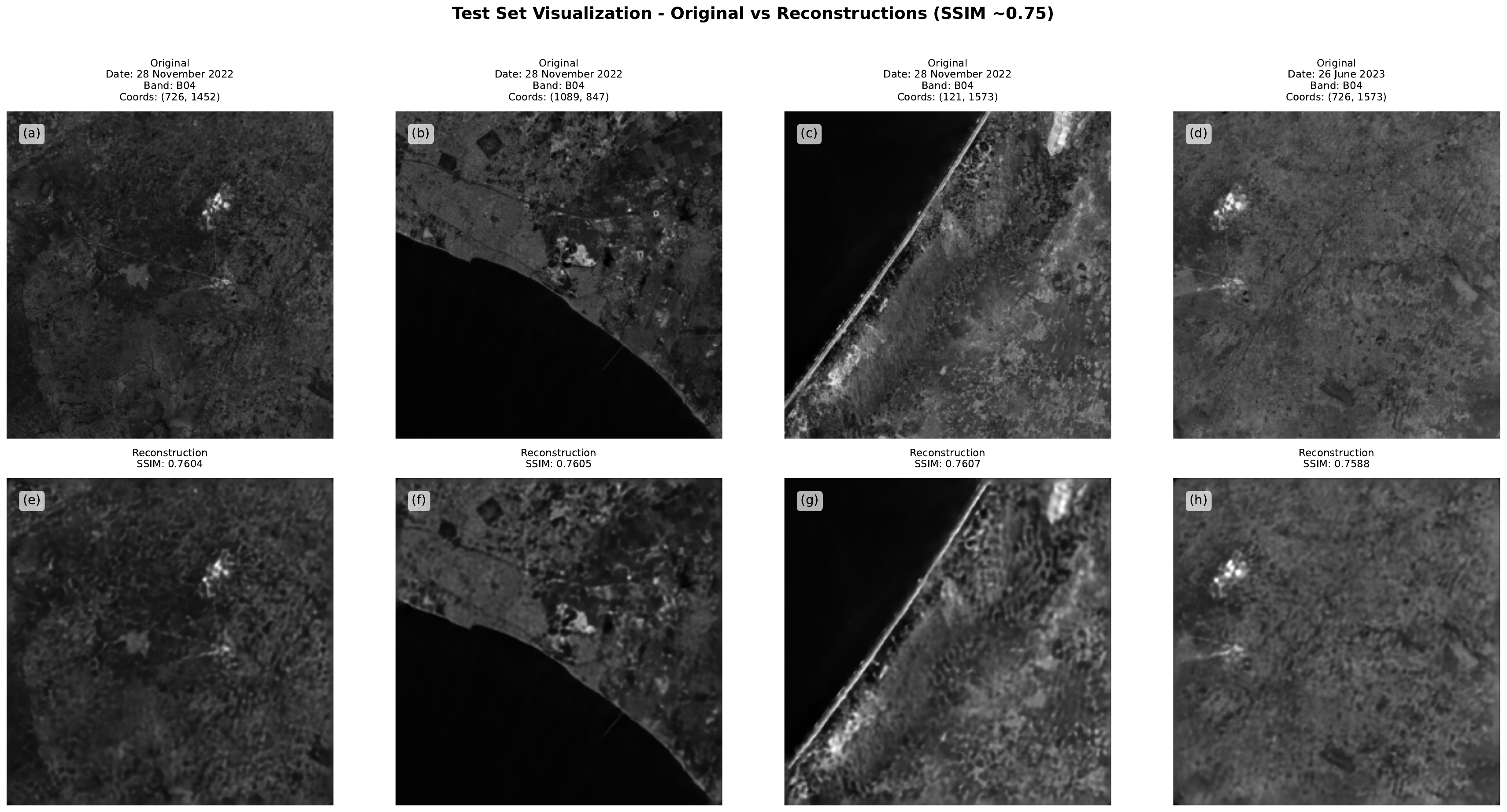}
\end{adjustwidth}
    \caption{Original test 
 set (\textbf{top}) and CAE-reconstructed (\textbf{bottom}) images with metadata to study the CAE performance with values around a mean test set SSIM score of 0.75. (a) Original test set image patch from 28 November 2022, Band B04, coordinates (726, 1452); (b) Original test set image patch from 28 November 2022, Band B04, coordinates (1089, 847); (c) Original test set image patch from 28 November 2022, Band B04, coordinates (121, 1573); (d) Original test set image patch from 26 June 2023, Band B04, coordinates (726, 1573); (e) CAE reconstruction of (a) with SSIM score of 0.7604; (f) CAE reconstruction of (b) with SSIM score of 0.7605; (g) CAE reconstruction of (c) with SSIM score of 0.7607; (h) CAE reconstruction of (d) with SSIM score of 0.7588.
    \label{fig:ex_recos}} 
\end{figure}

The images reconstructed by the CAE in the test set have a mean SSIM score of 0.75, indicating high-fidelity decoding by the model and validating the potential of transmitting images to Earth in a compressed format without the loss of key information. 
Figure~\ref{fig:ex_recos} shows some example reconstructions of the satellite image patches processed by the CAE decoder. 
Comparatively, the original result with this model structure in Ref.~\cite{GUERRISI2023} provides a mean test set SSIM score of 0.977. 
However, this is obtained by optimizing a model strictly for reconstruction capabilities, whereas high-quality reconstruction and anomaly detection are inherently competing tasks, thus limiting the optimal performance achievable for either one. 
The further results below provide a benchmark for the AD capability, obtained from a state-of-the-art reconstruction model, with potential for further targeted optimizations for the single-model deployment of both functionalities (as discussed in Section~\ref{subsec:disc}). 

\subsection{Anomaly Detection} 

\begin{figure}[H]

\begin{adjustwidth}{-\extralength}{0cm}
\centering 
 \includegraphics[width=0.95\linewidth]{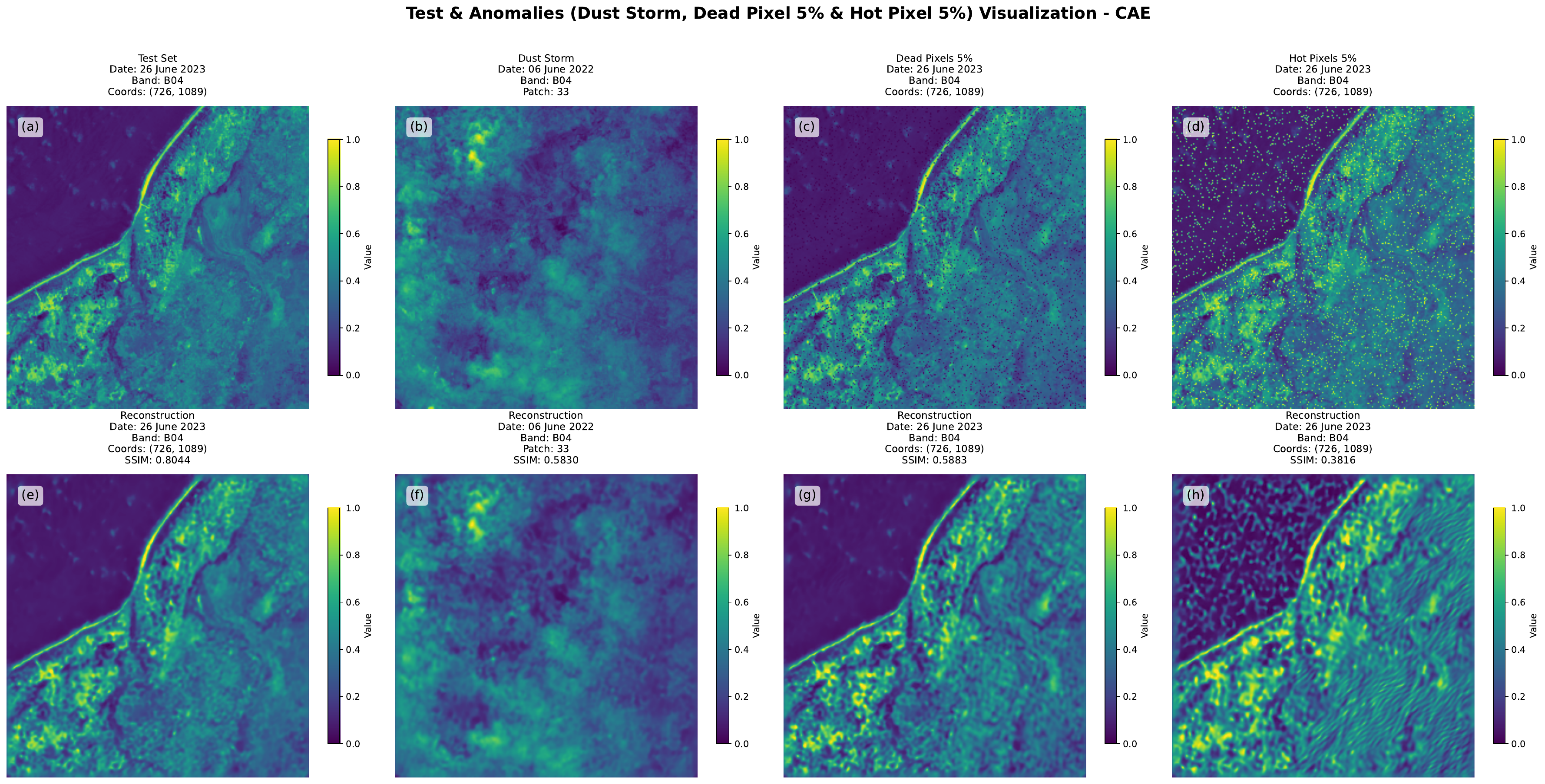}
\end{adjustwidth}
    \caption{True 
 (\textbf{top}) and CAE-reconstructed (\textbf{bottom}) images for the four test sets used to study the anomaly detection performance; from left to right, background, Sahara dust storm, dead pixel, and hot pixel images. Images are displayed in Viridis color scale for ease of visualization of anomalies. (a) True image patch from the normal test set, acquired on 26 June 2023, Band B04, coordinates (726, 1089); (b) True image patch from the dust storm anomaly test set, acquired on 06 June 2022, Band B04, patch 33; (c) True image patch with synthetic dead pixel anomaly (5\% of pixels set to 0), based on the 26 June 2023 image, Band B04, coordinates (726, 1089); (d) True image patch with synthetic hot pixel anomaly (5\% of pixels set to 1), based on the 26 June 2023 image, Band B04, coordinates (726, 1089); (e) CAE reconstruction of (a) with SSIM score of 0.8044; (f) CAE reconstruction of (b) with SSIM score of 0.5830; (g) CAE reconstruction of (c) with SSIM score of 0.5883; (h) CAE reconstruction of (d) with SSIM score of 0.3816.
    \label{fig:anomalies}} 
\end{figure}

In order to test the capability of the CAE to provide model-independent recognition of anomalous satellite images, a variety of both real and synthetic anomalous images are used in the test set. 
The first anomalies used are the real satellite images collected during known periods of Sahara dust storms, resulting in occluded images with higher-than-normal mean pixel values. 
Two other anomalies are created synthetically to mimic the impact of camera defects, namely, a ``dead pixel'' test set, where 5\% of the pixels in each patch have values set to 0, and a ``hot pixel'' test set, where 5\% of the pixels per patch are set to 1. 
Figure~\ref{fig:anomalies} provides examples of the three anomaly image classes compared to a normal background image, showing both the true image and the output image reconstructed by the CAE. 

\begin{figure}[H]

\begin{adjustwidth}{-\extralength}{0cm}
\centering 
 \includegraphics[width=0.45\linewidth]{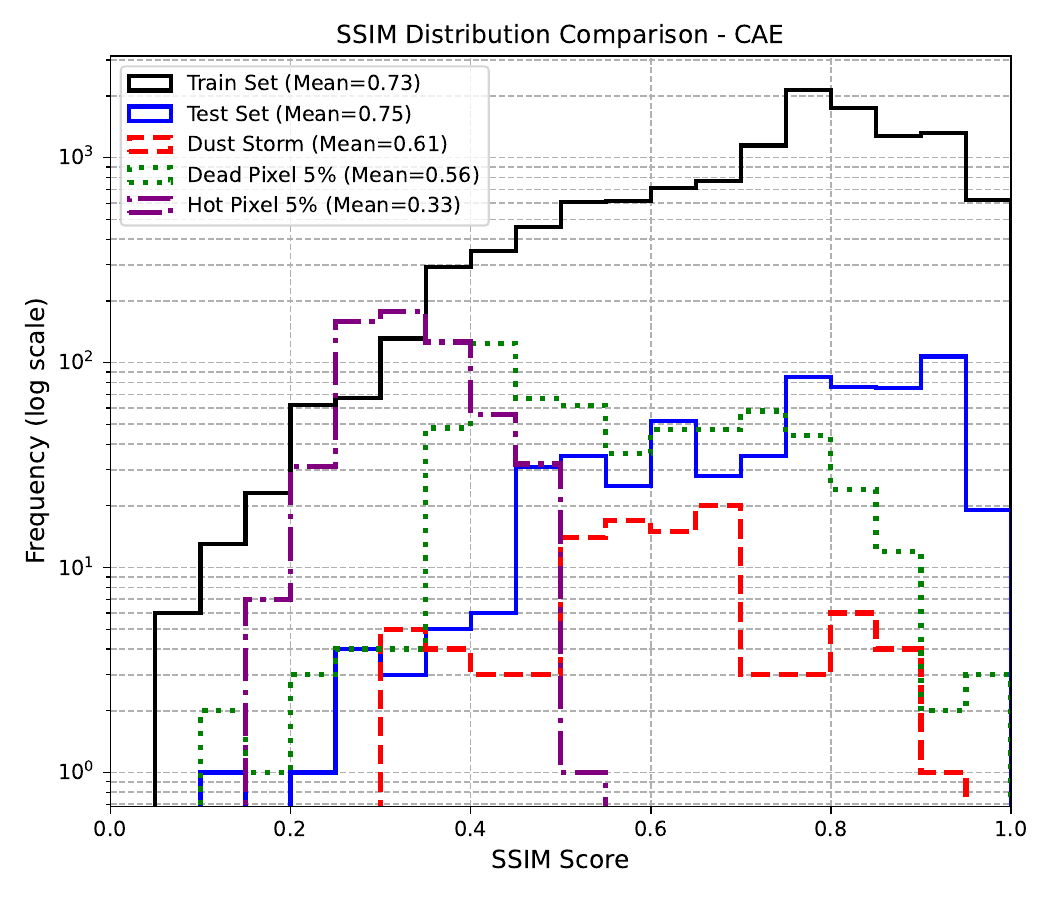}
    \includegraphics[width=0.5\linewidth]{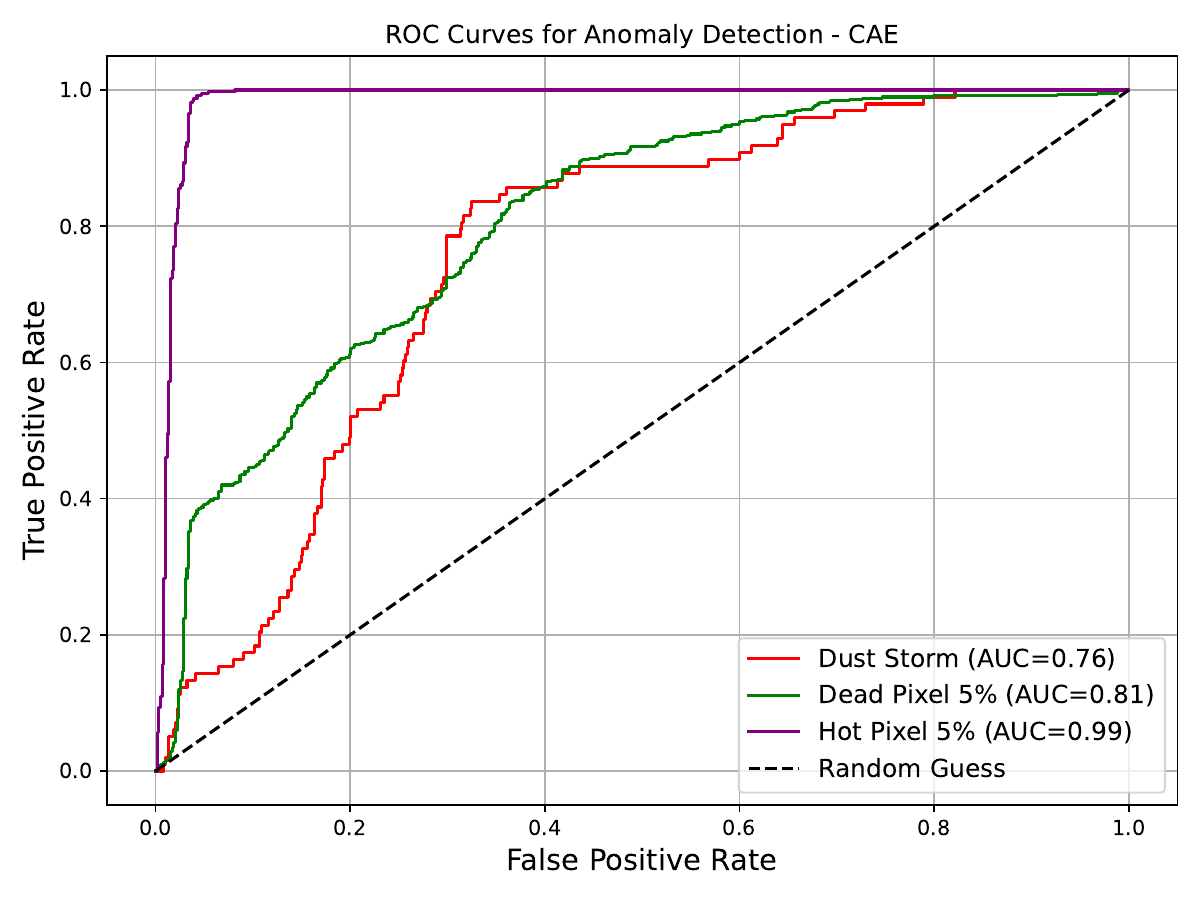}
\end{adjustwidth}
    \caption{Histogram of the CAE output SSIM score on the training, test, and three anomaly test sets (\textbf{left}) and the corresponding ROC curve (\textbf{right}). Higher SSIM values indicate better \mbox{reconstruction quality}. 
    \label{fig:ssim}} 
\end{figure}

Separate test sets are constructed comprising pure samples of each of these synthetic anomalies in order to assess the ability of the CAE SSIM score to isolate them from the background. 
Figure~\ref{fig:ssim} shows the resulting SSIM score distribution for the training, test, and three anomalous image sets, as well as the receiver operating characteristic (ROC) considering the SSIM as a classifier.
The anomaly test sets all share a lower mean SSIM score than the normal data, as represented by the training and test sets, with the hot pixel anomaly set appearing to have the lowest reconstruction quality and therefore the highest likelihood of anomalous origin. 
Each of the anomaly test sets is also identifiable via thresholding on the SSIM score, with ROC areas under the curves (AUCs) of 0.76 for the dust storm images, 0.81 for the dead pixel set, and 0.99 for the hot pixel set. 


\subsection{Discussion}
\label{subsec:disc}

Designing a data acquisition system for SmallSat that includes a CAE on the satellite could enable the recognition of anomalous images in real time and at the source of the data. 
This functionality could enable the rapid monitoring of procedural changes, such as the disruption of a standard satellite scanning process to remain on a particular aerial region or to take more detailed images.
In the context of the use case of Sahara dust storm identification, this capability can enhance real-time anomaly monitoring, leading to substantial potential impacts on disaster response and research. 


As discussed above, this work extends the CAE developed for small satellite image compression in Ref.~\cite{GUERRISI2023} and demonstrates that a single model can achieve both compression for off-satellite transmission and at-source anomaly detection, capable of being used to identify natural disasters or camera defects. 
Similar studies with deep ML models for image compression have been performed in small satellite environments~\cite{9172536, rs11070759, rs13030447, 9690871}, but none of them have explored the anomaly detection potential of these models. 

This work aims to start from an existing CAE demonstrated to be close to the state of the art in previous works.
However, further optimization of the CAE model could achieve gains in the image reconstruction quality, anomaly detection, or both. 
A study is conducted with this CAE to scan the model latent space size, considering alternate CR values of 4, 16, and 32. 
Compared to a CR of 8, which yields a mean SSIM score on the test set of 0.75, the higher CR values considered degrade image reconstruction quality significantly, with mean SSIM test set values of 0.66 and 0.51 for 16 and 32, respectively.
This indicates that significantly smaller latent spaces prevent the model from saving enough salient information to perform high-quality decompression, limiting the utility of the model for off-satellite data compression. 
The model with a lower CR of 4 has $\mathcal{O}$(1)\% level improvements in both the average image SSIM score and anomaly AUC. 
However, a larger latent space corresponds to reduced data compression and a larger volume to be transmitted.
Further optimization of the CR would require the consideration of the satellite payload constraints and data \mbox{acquisition needs. }

While the single-CAE approach is promising in its performance, as demonstrated here, and appealing in its simplicity for data acquisition designs, it suffers from some limitations:
the lossy nature of CAE-based compression means that the final dataset will have some imperfections with respect to the truth images. 
Careful estimation of this effect could be applied in downstream analysis as a systematic uncertainty, but this may not be acceptable for certain satellite applications.
Additionally, the inherent variability of Earth observation data could result in performance fluctuations or drift from a model trained on a fixed dataset, requiring careful monitoring of the algorithm and re-training or updating over time as needed. 
In light of these examples and other similar challenges, the adaptation of this concept for a specific data-taking scenario is key for deployment. 

This work provides a small-scale and simple proof of concept that modern AI-based image compression algorithms in use for SmallSat readout can also be used for anomaly detection in acquired images.
However, considerable future work could enable better performance on both tasks and bring this capability closer to deployment in real SmallSat payload compute hardware. 
This model is shown to recognize the common classes of anomalies that may appear in SmallSat images, namely, one natural disaster class and two camera errors.
The next step will be to explore the extrapolation of this model to recognize (a) anomalies of other kinds (e.g., flooding and volcanoes) and (b) different geographical regions beyond Dakar. 
Additional studies of the choice of architecture could improve the overall image reconstruction quality, for example, through the use of adversaries such as generative adversarial networks to fine-tune the CAE latent space and enhance output image fidelity. 
Finally, studies of the possible hardware deployment of this model with more realistic spectral instrumentation are needed, with potential exploration for deployment on traditional CPUs, as well as more efficient platforms such as field-programmable gate arrays (FPGAs).


\section{Conclusions}
\label{sec:conclusions}

The application of machine learning for small satellite data acquisition is presented for the use case of disaster monitoring in Africa.
Open-source datasets containing images taken over Dakar, Senegal, on the Sentinel-2 mission are used to develop a convolutional autoencoder with pixel-based input modeling. 
The resulting CAE loss upon test set evaluation is able to provide good reconstruction quality of the images, indicating the utility of on-board ML models to perform data compression at the edge to transmit lower data rates to Earth.
Additionally, the autoencoder output can identify anomalous events, such as those arising from Sahara dust storm events over Dakar or synthetic pixel issues.
Together, these capabilities motivate the use of ML-based methods in the design of on-satellite data acquisition systems for small satellites. 
The relatively low cost and accessibility of SmallSat technologies make these applications especially relevant for enhancing science and space programs in Africa while providing early career training opportunities in critical technologies. 


\vspace{6pt}

\authorcontributions{Conceptualization, J.G.; Methodology, D.J. and J.G.; Software, D.J.; Formal analysis, D.J.; Data curation, D.J.; Writing---original draft, D.J. and J.G.; Writing---review \& editing, J.G.; Supervision, J.G. All authors have read and agreed to the published version of the manuscript.}

\funding{This research was funded by U.S. Department of Energy under contract number DE-AC02-76SF00515.}

\institutionalreview{Not applicable.}


\informedconsent{Not applicable.}


\dataavailability{
The exact Sentinel-2 image set taken from Ref.~\cite{dataspace2024} and used in this study can be found on Zenodo~{\url{https://doi.org/10.5281/zenodo.15554086}
}.}

\acknowledgments{
The authors acknowledge the support of the Google DeepMind Scholarship Program at the African Institute for Mathematical Sciences (AIMS) South Africa, which enabled DJ to pursue the AI for Science Master’s program during the conduct of this research. This program provided essential academic training, computational resources, and mentorship that contributed directly to the development of the convolutional autoencoder model and the analysis of satellite imagery presented in this work.

}

\conflictsofinterest{The funders had no role in the design of the study; in the collection, analyses, or interpretation of data; in the writing of the manuscript, or in the decision to publish the~results.}


\begin{adjustwidth}{-\extralength}{0cm}

\reftitle{References}

\PublishersNote{}
\end{adjustwidth}

\begin{thebibliography}{999}

\end{thebibliography}


\begin{thebibliography}{999}

\bibitem[Joseph R.~Kopacz and Roney(2020)]{KOPACZ2020}
Joseph, R.;~Kopacz, R.H.; Roney, J.
\newblock {Small satellites an overview and assessment}.
\newblock {\em Acta Astronaut.} {\bf 2020}, {\em 170},~93--105.
\newblock
  {\url{https://doi.org/10.1016/j.actaastro.2020.01.034}}.

\bibitem[Fevgas et~al.(2025)Fevgas, Lagkas, Sarigiannidis, and
  Argyriou]{Fevgas2025}
Fevgas, G.; Lagkas, T.; Sarigiannidis, P.; Argyriou, V.
\newblock Advances in Remote Sensing and Propulsion Systems for Earth
  Observation Nanosatellites.
\newblock {\em Future Internet} {\bf 2025}, {\em 17},~16.
\newblock {\url{https://doi.org/10.3390/fi17010016}}.

\bibitem[Alzubairi et~al.(2024)Alzubairi, Tameem, and
  Kada]{ALZUBAIRI2024e02391}
Alzubairi, A.; Tameem, A.; Kada, B.
\newblock Spacecraft formation flying orbital control for earth observation
  mission.
\newblock {\em Sci. Afr.} {\bf 2024}, {\em 26},~e02391.
\newblock {\url{https://doi.org/10.1016/j.sciaf.2024.e02391}}.

\bibitem[Battistini(2022)]{BATTISTINI2022231}
Battistini, S.
\newblock Chapter 12-Small satellites for disaster monitoring. In {\em
  Nanotechnology-Based Smart Remote Sensing Networks for Disaster Prevention};
  Denizli, A., Alencar, M.S., Nguyen, T.A., Motaung, D.E., Eds.; Micro and Nano
  Technologies; Elsevier:  Amsterdam, The Netherlands, 
  2022; pp. 231--251.
\newblock
  {\url{https://doi.org/10.1016/B978-0-323-91166-5.00002-1}}.

\bibitem[Francesco~Barato and Pavarin(2024)]{BARATO2024}
Barato, F.; Toson, E.; Milza, F.; Pavarin, D.
\newblock {Investigation of different strategies for access to space of small
  satellites on a defined LEO orbit}.
\newblock {\em Acta Astronaut.} {\bf 2024}, 222, 11--28.
\newblock
  {\url{https://doi.org/10.1016/j.actaastro.2024.05.045}}.

\bibitem[Kulu(2024)]{KULU2024}
Kulu, E.
\newblock CubeSats \& Nanosatellites--2024 Statistics, Forecast and
  Reliability.
\newblock In Proceedings of the 75th International Astronautical Congress (IAC
  2024), International Astronautical Federation (IAF), Milan, Italy, 14--18 
 October
  2024; p. IAC--24.B4.6A.13. 
  


\bibitem[Yansheng~Li and Zhang(2021)]{LI2021}
Li, Y.; Ma, J.; Zhang, Y.
\newblock Image retrieval from remote sensing big data: A survey.
\newblock {\em Inf. Fusion} {\bf 2021}, \emph{67}, 94--115.
\newblock {\url{https://doi.org/10.1016/j.inffus.2020.10.008}}.

\bibitem[Chintalapati et~al.(2024)Chintalapati, Precht, Hanra, Laufer, Liwicki,
  and Eickhoff]{CHINTALAPATI2024}
Chintalapati, B.; Precht, A.; Hanra, S.; Laufer, R.; Liwicki, M.; Eickhoff, J.
\newblock {Opportunities and challenges of on-board AI-based image recognition
  for small satellite Earth observation missions}.
\newblock {\em Adv. Space Res.} {\bf 2024}, \emph{75}, 6734--6751.
\newblock {\url{https://doi.org/10.1016/j.asr.2024.03.053}}.

\bibitem[Alex et~al.(2012)Alex, Ilya, and Geoffrey]{Krizhevsky2012}
Krizhevsky, A.; Sutskever, I.; Hinton, G.E.
\newblock ImageNet Classification with Deep Convolutional Neural Networks.
\newblock In Proceedings of the Advances in Neural Information Processing
  Systems 25 (NIPS 2012),  Lake Tahoe, NV, USA, 3–6 December 2012; Volume~25, pp. 1097--1105.

\bibitem[Lofqvist and Cano(2020)]{lofqvist2020}
Lofqvist, M.; Cano, J.
\newblock Accelerating Deep Learning Applications in Space. 
\newblock  \emph{arXiv} \textbf{2020},  arXiv:2007.11089. 

\bibitem[Chen et~al.(2025)Chen, Gao, Lu, Zhang, Ding, Li, and
  Zhang]{CHEN2025104106}
Chen, Z.; Gao, H.; Lu, Z.; Zhang, Y.; Ding, Y.; Li, X.; Zhang, B.
\newblock MDA-HTD: Mask-driven dual autoencoders meet hyperspectral target
  detection.
\newblock {\em Inf. Process. Manag.} {\bf 2025}, {\em
  62},~104106.
\newblock {\url{https://doi.org/10.1016/j.ipm.2025.104106}}.

\bibitem[Gianluca et~al.(2021)Gianluca, Luca, Gabriele, Matej, L\'{e}onie, and
  Aubrey]{GIUFFRIDA2021}
Gianluca, G.; Luca, F.; Gabriele, M.; Matej, B.; L\'{e}onie, B.; Aubrey, D.
\newblock The $\Phi$-Sat-1 Mission: The First On-Board Deep Neural Network
  Demonstrator for Satellite Earth Observation.
\newblock {\em IEEE Trans. Geosci. Remote Sens.} {\bf 2021},
  {\em 60},~5517414.
\newblock {\url{https://doi.org/10.1109/TGRS.2021.3125567}}.

\bibitem[Goodwill et~al.(2020)Goodwill, Wilson, Sabogal, George, and
  Wilson]{9172536}
Goodwill, J.; Wilson, D.; Sabogal, S.; George, A.D.; Wilson, C.
\newblock Adaptively Lossy Image Compression for Onboard Processing.
\newblock In Proceedings of the 2020 IEEE Aerospace Conference,  Big Sky, MT, USA, 7--14 March 2020; pp.
  1--15.
\newblock {\url{https://doi.org/10.1109/AERO47225.2020.9172536}}.

\bibitem[Li and Liu(2019)]{rs11070759}
Li, J.; Liu, Z.
\newblock Multispectral Transforms Using Convolution Neural Networks for Remote
  Sensing Multispectral Image Compression.
\newblock {\em Remote Sens.} {\bf 2019}, {\em 11}, 759.
\newblock {\url{https://doi.org/10.3390/rs11070759}}.

\bibitem[Alves~de Oliveira et~al.(2021)Alves~de Oliveira, Chabert, Oberlin,
  Poulliat, Bruno, Latry, Carlavan, Henrot, Falzon, and Camarero]{rs13030447}
Alves~de Oliveira, V.; Chabert, M.; Oberlin, T.; Poulliat, C.; Bruno, M.;
  Latry, C.; Carlavan, M.; Henrot, S.; Falzon, F.; Camarero, R.
\newblock Reduced-Complexity End-to-End Variational Autoencoder for on Board
  Satellite Image Compression.
\newblock {\em Remote Sens.} {\bf 2021}, \mbox{{\em 13}, 447}.
\newblock {\url{https://doi.org/10.3390/rs13030447}}.

\bibitem[Alves~de Oliveira et~al.(2022)Alves~de Oliveira, Chabert, Oberlin,
  Poulliat, Bruno, Latry, Carlavan, Henrot, Falzon, and Camarero]{9690871}
Alves~de Oliveira, V.; Chabert, M.; Oberlin, T.; Poulliat, C.; Bruno, M.;
  Latry, C.; Carlavan, M.; Henrot, S.; Falzon, F.; Camarero, R.
\newblock Satellite Image Compression and Denoising With Neural Networks.
\newblock {\em IEEE Geosci. Remote Sens. Lett.} {\bf 2022}, {\em
  19},~4504105.
\newblock {\url{https://doi.org/10.1109/LGRS.2022.3145992}}.

\bibitem[Paszkowsky et~al.(2020)Paszkowsky, Br\"{a}nnvall, Carlstedt, Milz,
  Kov\'{a}cs, and Liwicki]{AGUESPASZKOWSKY2020}
Paszkowsky, N.A.; Br\"{a}nnvall, R.; Carlstedt, J.; Milz, M.; Kov\'{a}cs, G.;
  Liwicki, M.
\newblock {Vegetation and Drought Trends in Sweden's M\"{a}lardalen Region
  \textendash{} Year-on-Year Comparison by Gaussian Process Regression}.
\newblock In Proceedings of the 2020 Swedish Workshop on Data Science (SweDS),
  Luleå, Sweden, 29--30 October 2020.
\newblock {\url{https://doi.org/10.1109/SweDS51247.2020.9275587}}.

\bibitem[Lin et~al.(2024)Lin, Zhang, Cheng, Shi, Gamba, and Wang]{10285414}
Lin, S.; Zhang, M.; Cheng, X.; Shi, L.; Gamba, P.; Wang, H.
\newblock Dynamic Low-Rank and Sparse Priors Constrained Deep Autoencoders for
  Hyperspectral Anomaly Detection.
\newblock {\em IEEE Trans. Instrum. Meas.} {\bf
  2024}, {\em 73},~2500518.
\newblock {\url{https://doi.org/10.1109/TIM.2023.3323997}}.

\bibitem[Bogdan~Ruszczak(2025)]{natureAD}
Bogdan~Ruszczak, Krzysztof~Kotowski, D.E.J.N.
\newblock {The OPS-SAT benchmark for detecting anomalies in satellite
  telemetry}.
\newblock {\em Sci. Data} {\bf 2025}, {\em 12},~710.
\newblock {\url{https://doi.org/10.1038/s41597-025-05035-3}}.

\bibitem[Liang~Liu(2023)]{telemetry2}
Liu, L.; Tian, L.; Kang, Z.; Wan, T.
\newblock {Spacecraft anomaly detection with attention temporal convolution
  networks}.
\newblock {\em Neural Comput. Appl.} {\bf 2023}, {\em 35},~9753--9761.
\newblock {\url{https://doi.org/10.1007/s00521-023-08213-9}}.

\bibitem[Hussein(2023)]{timeSeries}
Hussein, M.
\newblock A Real-Time Anomaly Detection in Satellite Telemetry Data Using
  Artificial Intelligence Techniques Depending on Time-Series Analysis.
\newblock {\em J. ACS Adv. Comput. Sci.} {\bf 2023},
  {\em 14}(1), 21--45. 

  
\newblock {\url{https://doi.org/10.21608/asc.2023.171575.1011}}.

\bibitem[Guerrisi et~al.(2022)Guerrisi, Del~Frate, and Schiavon]{9883256}
Guerrisi, G.; Del~Frate, F.; Schiavon, G.
\newblock {Convolutional Autoencoder Algorithm for On-Board Image Compression}.
\newblock In Proceedings of the IGARSS 2022-2022 IEEE International
  Geoscience and Remote Sensing Symposium, Kuala Lumpur, Malaysia, 17--22 July 2022; pp. 151--154.
\newblock {\url{https://doi.org/10.1109/IGARSS46834.2022.9883256}}.

\bibitem[Daniel and David(2012)]{SELVA2012}
Daniel, S.; David, K.
\newblock A survey and assessment of the capabilities of Cubesats for Earth
  observation.
\newblock {\em Acta Astronaut.} {\bf 2012}, {\em 74},~50--68.
\newblock
  {\url{https://doi.org/10.1016/j.actaastro.2011.12.014}}.

\bibitem[{Copernicus Programme}(2024)]{dataspace2024}
{Copernicus Programme}.
\newblock {Copernicus Data Space Ecosystem: Sentinel-2 Data Collection}.
\newblock 2024.
\newblock
  Available online: \url{https://dataspace.copernicus.eu/explore-data/data-collections/sentinel-data/sentinel-2} (accessed on 21 November 2024).


\bibitem[Giuliano et~al.(2024)Giuliano, Gadsden, Hilal, and
  Yawney]{giuliano2024}
Giuliano, A.; Gadsden, S.A.; Hilal, W.; Yawney, J.
\newblock Convolutional variational autoencoders for secure lossy image
  compression in remote sensing.  \emph{arXiv} \textbf{2024}, arXiv:2404.03696.

\bibitem[O'Sullivan et~al.(2020)O'Sullivan, Marenco, Ryder, Pradhan, Kipling,
  Johnson, Benedetti, Brooks, McGill, Yorks, and Selmer]{O'Sullivan2020}
O'Sullivan, D.; Marenco, F.; Ryder, C.L.; Pradhan, Y.; Kipling, Z.; Johnson,
  B.; Benedetti, A.; Brooks, M.; McGill, M.; Yorks, J.;  et~al.
\newblock Models transport Saharan dust too low in the atmosphere: A comparison
  of the MetUM and CAMS forecasts with observations.
\newblock {\em Atmos. Chem. Phys.} {\bf 2020}, {\em
  20},~12955--12982.
\newblock {\url{https://doi.org/10.5194/acp-20-12955-2020}}.


 

\bibitem[Phiri et~al.(2020)Phiri, Simwanda, Salekin, Nyirenda, Murayama, and
  Ranagalage]{rs12142291}
Phiri, D.; Simwanda, M.; Salekin, S.; Nyirenda, V.R.; Murayama, Y.; Ranagalage,
  M.
\newblock Sentinel-2 Data for Land Cover/Use Mapping: A Review.
\newblock {\em Remote Sens.} {\bf 2020}, {\em 12}, 2291.
\newblock {\url{https://doi.org/10.3390/rs12142291}}.



\bibitem[{NASA Earth Observatory}(2022)]{duststorm}
{NASA Earth Observatory}.
\newblock A Burst of Saharan Dust,  2022. 
\newblock  Online resource: \url{https://earthobservatory.nasa.gov/images/149918/a-burst-of-saharan-dust}. 
Accessed: 05 August 2024.




\bibitem[Guerrisi et~al.(2023)Guerrisi, Frate, and Schiavon]{GUERRISI2023}
Guerrisi, G.; Frate, F.D.; Schiavon, G.
\newblock {Artificial Intelligence Based On-Board Image Compression for the
  $\phi$-Sat-2 Mission}.
\newblock {\em IEEE J. Sel. Top. Appl. Earth Obs. Remote Sens.} {\bf 2023}, {\em 16},~8063--8075.
\newblock {\url{https://doi.org/10.1109/JSTARS.2023.3296485}}.

\bibitem[Dosselmann and Yang(2011)]{Dosselmann2011}
Dosselmann, R.; Yang, X.D.
\newblock A comprehensive assessment of the structural similarity index.
\newblock {\em Signal Image Video Process.} {\bf 2011}, {\em 5},~81--91.
\newblock {\url{https://doi.org/10.1007/s11760-009-0144-1}}.

\bibitem[Horé and Ziou(2010)]{HORE2010}
Horé, A.; Ziou, D.
\newblock Image quality metrics: PSNR vs. SSIM. In Proceedings of the 2010 20th International Conference on Pattern Recognition,  Istanbul, Turkey, 23--26 August 2010; 
\newblock  pp. 2366--2369. \newblock {\url{https://doi.org/10.1109/ICPR.2010.579}}.

\bibitem[Kingma and Ba(2017)]{kingma2017adammethodstochasticoptimization}
Kingma, D.P.; Ba, J.
\newblock Adam: A Method for Stochastic Optimization. \emph{arXiv}  \textbf{2017}, arXiv:1412.6980.

\end{thebibliography}
\end{document}